\documentstyle[12pt]{article}
\input{epsf.tex}

\begin{document}

  \begin{flushright} \begin{small}
  DF/IST-4.2001 \\gr-qc/0105103
  \end{small} \end{flushright}

\vskip 0.5cm

\begin{center}
{\bf QUASI-NORMAL MODES \\
OF SCHWARZSCHILD ANTI-DE SITTER BLACK HOLES:\\
ELECTROMAGNETIC AND GRAVITATIONAL PERTURBATIONS
} \\
\vskip 0.5cm
Vitor Cardoso\footnote{E-mail: vcardoso@fisica.ist.utl.pt} \\
\vskip 0.2cm
{\scriptsize  CENTRA, Departamento de F\'{\i}sica,
	      Instituto Superior T\'ecnico,} \\

{\scriptsize  Av. Rovisco Pais 1, 1096 Lisboa, Portugal.} \\
\vskip 0.5cm

Jos\'e P. S. Lemos\footnote{E-mail: lemos@kelvin.ist.utl.pt} \\
\vskip 0.2cm
{\scriptsize  CENTRA, Departamento de F\'{\i}sica,
	      Instituto Superior T\'ecnico,} \\
{\scriptsize  Av. Rovisco Pais 1, 1096 Lisboa, Portugal.}
\end{center} 
 
\vskip 0.1cm

\begin{abstract}
\noindent
We study the quasi-normal modes (QNM) of electromagnetic and
gravitational perturbations of a Schwarzschild black hole in an
asymptotically Anti-de Sitter (AdS) spacetime.  Some of the
electromagnetic modes do not oscillate, they only decay, since they
have pure imaginary frequencies.  The gravitational modes show
peculiar features: the odd and even gravitational perturbations no
longer have the same characteristic quasinormal frequencies.  There is
a special mode for odd perturbations whose behavior differs completely
from the usual one in scalar and electromagnetic perturbation in an
AdS spacetime, but has a similar behavior to the Schwarzschild black
hole in an asymptotically flat spacetime: the imaginary part of the
frequency goes as $\frac{1}{r_+}$, where $r_+$ is the horizon radius.
We also investigate the small black hole limit showing that the
imaginary part of the frequency goes as $r_+^2$. These results are
important to the AdS/CFT conjecture since according to it the QNMs
describe the approach to equilibrium in the conformal field theory.
\strut  
\newline 
\end{abstract}

\noindent
\section{ Introduction}
\vskip 3mm
QNMs of black holes play an important role in the study of the
dynamics outside black holes. They appear, for instance, when one
deals with the evolution of some field in the black hole spacetime, or
in black hole-black hole collision processes. Numerical simulations
ranging from the formation of a black hole in a gravitational collapse
\cite{Cunningham} to the collision of two black
holes \cite{Anninos} provide clear evidence that
no matter how one perturbs a black hole, its response will be
dominated by the QNMs.  QNMs allow us not only to test the stability
of the event horizon against small perturbations, but also to probe
the black hole mass, electric charge and angular momentum, through
their characteristic waveform.

A great deal of effort has been spent to calculate the QNMs and their
associated frequencies. New powerful methods, both analytical and
numerical have been developed. The main interest in these studies  
is in the application to the analysis of the data from the gravitational 
waves to be detected by the forthcoming  gravitational wave
detectors. We refer the reader to \cite{Kokkotas,Andersson} for 
reviews.  In a different context, York \cite{York} tried to explain the
thermal quantum radiance of a Schwarzschild black hole in terms of 
quantum zero-point fluctuations of zero mean in the QNMs.

All these previous works deal with
asymptotically flat spacetimes, but the recent 
AdS/CFT correspondence conjecture \cite{maldacena}
makes the investigation of QNMs in anti-de 
Sitter spacetimes more appealing. According to it, the black hole
corresponds to a thermal state in the conformal field theory, and the decay of
the test field in the black hole spacetime, corresponds to the decay of the perturbed state
in the CFT.
The dynamical timescale for the return to thermal equilibrium is very
hard to compute directly, but can be done relatively easily using the
AdS/CFT correspondence. Horowitz and Hubeny
\cite{Horo} (see also \cite{Horowitz}) began the study of QNMs 
in AdS, by thoroughly investigating scalar perturbations in 4, 5
and 7 spacetime dimensions. Subsequently, Wang and et al
\cite{Wang1,Wang2} analyzed scalar QNMs in a Reissner-Nordst\"om AdS
geometry. Recently, Cardoso and Lemos \cite{Cardoso} found an exact
solution for the QNMs of scalar, electromagnetic and Weyl
perturbations of a BTZ black hole. 
Another conjecture is related to the speculation \cite{Horo,Bir,Kim} that
there might be a connection between the critical exponent of Choptuik
\cite{Chop} and the imaginary part of the frequency, for small black
holes. This is still an open question.

In this paper we shall go beyond the scalar perturbations
\cite{Horo,Wang1,Wang2}, and consider electromagnetic and
gravitational perturbations of a Schwarzschild black hole in an
asymptotically AdS spacetime.
Electromagnetic
perturbations are of interest due to the AdS/CFT conjecture since they can be seen
as perturbations for some generic supergravity gauge field.
In addition, the Maxwell field is an important field with different features
from scalar or gravitational fields, which makes it worth studying. 
On the other hand,  gravitational perturbations
have the additional interest of arising from any other type of
perturbation, be it scalar, electromagnetic, Weyl, etc., which in turn
disturb the background geometry. Therefore, questions like the
stability of spacetime for scalar or other perturbations, have a
direct dependence on the stability to gravitational perturbations.

We will find that in the case of electromagnetic perturbations of
large black holes, the characteristic QNM frequencies have only an
imaginary part, and scale with the horizon radius.  As for
gravitational perturbations, there are two novel features. First,
contrary to the asymptotically flat spacetime case, odd and even
perturbations no longer have the same spectra, although in certain
limits one can still prove that the frequencies are almost the same.
The second intriguing result is that, for odd perturbations, there is
a mode with a totally different behavior from that found in the
scalar and electromagnetic case: in this mode the frequency scales
with $\frac{1}{r_+}$, just as in asymptotically flat Schwarzschild
spacetime.  We also investigate the small black hole limit (a problem
recently addressed by Zhu et al \cite{Wang3}), and find that the QNM
frequencies go as $r_+^2$.


\noindent
\section{Electromagnetic and Gravitational perturbations in a 
Schwarzschild AdS background}


\subsection{Maxwell perturbations}
We consider the evolution of a Maxwell field in a
Schwarzschild-anti-de Sitter spacetime with metric given by 
\begin{equation}
ds^{2}= f(r) dt^{2}- \frac{dr^{2}}{f(r)}-
r^{2}(d\theta^{2}+\sin^2\theta d\phi^{2})\,,
\label{lineelement}
\end{equation}
where, $f(r)=(\frac{ r^{2}}{R^2}+1-\frac{2M}{r})$,
$R$ is the AdS radius and $M$ the black hole mass.  
The evolution is governed by Maxwell's equations:
\begin{equation}
{F^{\mu\nu}}_{;\nu}=0 \quad, F_{\mu\nu}=A_{\nu,\mu}-A_{\mu,\nu}\,,
\label{maxwell} 
\end{equation}
where a comma stands for ordinary derivative and a semi-colon 
for covariant derivative.  As the background is spherically symmetric,
we can expand $A_{\mu}$ in 4-dimensional vector sphericall
harmonics (see \cite{Ruffini}):

{\small
\begin{eqnarray}
A_{\mu}(t,r,\theta,\phi)=\sum_{l,m}\left( \begin{array}{cc}\left[
 \begin{array}{c} 0 \\ 0 \\
 \frac{a^{lm}(t,r)}{\sin\theta}\partial_\phi Y_{lm}\\
 -a^{lm}(t,r)\sin\theta\partial_\theta Y_{lm}\end{array}\right] &
 +\left[ \begin{array}{c}f^{lm}(t,r)Y_{lm}\\h^{lm}(t,r)Y_{lm} \\
 k^{lm}(t,r) \partial_\theta Y_{lm}\\ k^{lm}(t,r) \partial_\phi
 Y_{lm}\end{array}\right] \end{array}\right)\,,
\label{expansion}
\end{eqnarray}}

\noindent where the first term in the right-hand side has parity $(-1)^{l+1}$
and the second term has parity $(-1)^{l}$, $m$ is the azimuthal number
and $l$ the angular quantum number.  If we put this expansion into Maxwell's
equations (\ref{maxwell}) we get a second order differential
equation for the perturbation:
\begin{equation}
\frac{\partial^{2} \Psi(r)}{\partial r_*^{2}} +
\left\lbrack\omega^2-V(r)\right\rbrack\Psi(r)=0 \,,
\label{wavemaxwell}
\end{equation}
where the wavefunction $\Psi(r)$ is a linear combination of the functions
$f^{lm}$, $h^{lm}$, $k^{lm}$ and $a^{lm}$ as appearing in
(\ref{expansion}). $\Psi$ has a different functional dependence
according to the parity: for odd parity, i.e, $(-1)^{l+1}$, $\Psi$
is explicitly given by $\Psi=a^{lm}$ whereas for even parity $(-1)^l$
it is given by $\Psi=\frac{r^2}{l(l+1)}\left(-i\omega
h^{lm}-\frac{df^{lm}}{dr}\right)$, see \cite{Ruffini} for further details.
It is assumed that the time dependence is $\Psi(t,r)=e^{-i\omega t}\Psi(r)$.
The potential $V$ appearing in equation (\ref{wavemaxwell}) is given by
\begin{equation}
V(r)=f(r)\left\lbrack\frac{l(l+1)}{r^2}\right\rbrack \,,
\label{potentialmaxwell}
\end{equation}
and the tortoise coordinate $r_*$ is defined as
\begin{equation}
\frac{\partial r}{\partial r_*}= f(r)\,.
\end{equation}
We can of course rescale $r$, $r\rightarrow\frac{r}{R}$ and if we do
this, the wave equation again takes the form (\ref{wavemaxwell}) with
rescaled constants i.e., $r_+ \rightarrow \frac{r_+}{R}$, $\omega
\rightarrow \omega R $, where $r_+$ is the horizon radius.  So, we can
take $R=1$ and measure everything in terms of $R$.
\subsection{ Gravitational perturbations}
When dealing with first order gravitational perturbations one supposes
that, at least in some restricted region of spacetime, the metric
functions can be written as
\begin{equation}
g_{ab}(x^\nu)= g^{(0)}_{ab}(x^\nu)+h_{ab}(x^\nu)\,,
\label{4.1}
\end{equation}
where the metric $g^{(0)}_{ab}(x^\nu)$ is the background metric,
given by some known solution of Einstein's equations, and $ h_{ab}(x^\nu)$ is
a small perturbation.  Our background metric is a
Schwarszchild-anti-de Sitter metric (\ref{lineelement}) and the metric
$g_{ab}(x^\nu)$ will follow Einstein's equations in vacuum with a
cosmological constant:
\begin{equation}
G_{ab}-\Lambda g_{ab}=0\,.
\label{einstein}
\end{equation}
Upon substituting (\ref{4.1}) in (\ref{einstein}) we will obtain some
differential equations for the perturbations. We use the same
perturbations as originally given by Regge and Wheeler \cite{Regge},
retaining their notation. After a decomposition in tensorial spherical
harmonics (see Zerilli \cite{Zerilli1} and Mathews \cite{Mathews}),
these fall into two distinct classes - odd and even - with parities
$(-1)^{l+1}$ and $(-1)^l$ respectively, where $l$ is the angular
momentum of the particular mode.  While working in general relativity
one has some gauge freedom in choosing the elements $ h_{ab}(x^\nu)$
and one should take advantage of that freedom in order to simplify
the rather lengthy calculations involved in computing
(\ref{einstein}). We shall therefore work with the classical
Regge-Wheeler gauge in which the canonical form for the perturbations
is (see also \cite{Vish1}):

\bigskip
\noindent { odd \, parity:}
\begin{eqnarray}
h_{\mu \nu}= \left[
 \begin{array}{cccc} 
 0 & 0 &0 & h_0(r) 
\\ 0 & 0 &0 & h_1(r)
\\ 0 & 0 &0 & 0
\\ h_0(r) & h_1(r) &0 &h_0(r)
\end{array}\right] e^{-i \omega t}
\left(\sin\theta\frac{\partial}{\partial\theta}\right)
P_l(\cos\theta)\,;
\label{odd}
\end{eqnarray}
\bigskip
{ even \, parity:}
\begin{eqnarray}
h_{\mu \nu}= \left[
 \begin{array}{cccc} 
 H_0(r) f(r) & H_1(r) &0 & 0 
\\ H_1(r) & H_2(r)/f(r)  &0 & 0
\\ 0 & 0 &r^2K(r) & 0
\\ 0 & 0 &0 & r^2K(r)\sin^2\theta
\end{array}\right] e^{-i \omega t}
P_l(\cos\theta).
\label{even}
\end{eqnarray}
Here $P_l(\cos\theta)$ is the Legendre polynomial with angular
momentum $l$. 
If we put this decomposition into Einstein's equations we get ten
coupled second order differential equations that fully describe the
perturbations: three equations for odd perturbations and seven for
even perturbations. It is however possible to circumvent the
task of solving these coupled equations. Regge and Wheeler
\cite{Regge} and Zerilli \cite{Zerilli2} showed how to combine these
ten equations into two second order differential equations, one for
each parity. So following Regge and Wheeler \cite{Regge} (see also
\cite{Vish2} for more details) we define, for odd parity the wave 
function $Q(r)$ given by
perturbations,
\begin{equation}
Q(r)=  \frac{f(r)}{r} h_1(r) \,.
\label{qodd}
\end{equation}
After some work, Einstein's equations yield
\begin{equation}
\frac{\partial^{2} Q}{\partial r_*^{2}} +
\left\lbrack\omega^2 -V_{\rm odd}(r)\right\rbrack Q=0 \,,
\label{waveodd}
\end{equation}
where
\begin{equation}
V_{\rm odd}=  f(r)    
\left\lbrack\frac{l(l+1)}{r^2}-\frac{6m}{r^3}\right\rbrack\,.
\label{vodd}
\end{equation}
Likewise, following Zerilli \cite{Zerilli2} one can define for even modes the 
wavefunction $T(r)$ implicitly in terms of $H_0$, $H_1$ and $K$, through the 
equations
\begin{eqnarray}
K=
\frac{6m^2+c\left(1+c\right)r^2+m\left(3cr-3\frac{r^3}{R^2}\right)}
{r^2\left(3m+cr\right)} T
+\frac{dT}{dr_*} \,, \\
H_1= -\frac{i\omega\left(-3m^2-3cmr+cr^2-3m\frac{r^3}{R^2}\right)}
{r \left(3m+cr\right) f(r)} T -i \omega
\frac{r}{f(r)}\frac{dT}{dr_*}\,, 
\label{4.10}
\end{eqnarray}
where $c=\frac{1}{2}\left\lbrack l(l+1)-2\right\rbrack$.
Then Einstein's equations for even parity perturbations can be written as
\begin{equation}
\frac{\partial^{2} T}{\partial r_*^{2}} +
\left\lbrack \omega^2 -V_{\rm even}(r)\right\rbrack T=0 \,,
\label{waveeven}
\end{equation}
with
\begin{equation}
V_{\rm even}= \frac{2f(r)}{r^3}
\frac{9m^3+3c^2mr^2+c^2\left(1+c\right)r^3+3m^2\left(3cr+3\frac{r^3}{R^2}\right)}
{\left(3m+cr\right)^2} \,.
\label{veven}
\end{equation}
Now, by defining
\begin{equation} 
W=
\frac{2m}{r^2}+\frac{-3-2c}{3r}+\frac{3c^2+2c^2+27\frac{m^2}{R^2}}
{3c\left(3m+cr\right)}+j\,,
\label{W}
\end{equation}
where $j=-\frac{1}{3}\left(\frac{c}{m}+\frac{c^2}{m}+\frac{9m}{cR^2}\right)$, we obtain
\begin{equation}
V_{\rm odd}=W^2 {+} \frac{dW}{dr_*} +\beta \,, 
\quad 
V_{\rm even}=W^2 {-} \frac{dW}{dr_*} +\beta \,,
\label{V}
\end{equation}
where 
$\beta=-\frac{c^2+2c^3+c^4}{9m^2}$.  It is interesting to note that the two
potentials, odd and even, can be written in such a simple form, a fact
which seems to have been discovered by Chandrasekhar
\cite{Chandra2}. Potentials related in this manner are sometimes
called super-partner potentials \cite{Cooper}).  We note that similar
equations were obtained by Mellor and Moss \cite{Mellor} for
Schwarzschild-de Sitter spacetime, using a different approach.


\noindent
\section{Quasinormal modes and some of its properties}


\subsection{Analytical properties}
To solve equation (\ref{wavemaxwell}) for Maxwell fields 
and equations (\ref{waveodd}-\ref{waveeven}) for gravitational 
fields, one must
specify boundary conditions. 
Consider first the case of a Schwarzschild black hole in an
asymptotically flat spacetime (see, e.g., \cite{Kokkotas}). Since 
in this case the
potential  vanishes at both infinity and horizon, the two
solutions near these points are plane waves of the type 
$\Psi \sim e^{\pm i\omega r_*}$, where the $r_*$ coordinate
in this case ranges from $-\infty $ to $\infty$.  Quasinormal modes 
are defined by the condition that at the horizon there are only ingoing waves,
i.e., $\Psi_{\rm hor}\sim e^{-i\omega r_*} $ . Furthermore, one does not want to
have fields coming in from infinity (where the potential in this case 
vanishes). So, there is only a purely outgoing wave at infinity, i.e., 
$\Psi_{\rm \infty}\sim e^{i\omega r_*} $. Only a discrete set of complex 
frequencies $\omega$ meet these requirements.

Consider now a Schwarzschild black hole in an 
asymptotically AdS spacetime.  The boundary condition at the horizon 
is the same, we want that near the horizon             
$\Psi_{\rm hor}\sim e^{-i\omega r_*} $. However, $r_*$ has a finite range,
so the second boundary condition needs to be modified. There have been
several papers discussing which boundary conditions one should impose at infinity
in AdS spacetimes (\cite{Avis,Breit,Burgess}).  We
shall require energy conservation and thus adopt the reflective boundary
conditions at infinity \cite{Avis}.
This means that the wavefunction is zero at infinity. For 
a different boundary condition see \cite{Dasgupta}.

We now show that the imaginary part of the frequency $\omega$ is negative, for 
waves satisfying these boundary conditions, provided the potential $V$ is positive.
The proof proceeds as for the scalar field perturbation case  \cite{Horo}, 
although there are some steps we think are useful to display explicitly here. 
Writing $\phi$ for a generic wavefunction as 
\begin{equation}
\phi=e^{i\omega r_*} Z\,, 
\label{generic}
\end{equation}
where, $Z$ can be $\Psi$, $Q$ or $T$, we find 
\begin{equation}
f(r)\frac{\partial^{2} \phi}{\partial r^{2}}+
\left\lbrack f'-2i\omega \right\rbrack
\frac{\partial \phi}{\partial r} -\frac{V}{f}\phi=0\,,
\label{waveeq}
\end{equation}
where $f=( r^{2}+1-\frac{2M}{r})$.
In the proof, we are going to need the asymptotic behavior of
the solutions of equation (\ref{waveeq}).
For $r \rightarrow r_+$ we have $f \sim (3r_+ +\frac{1}{r_+})(r-r_+)$
and $\frac{V}{f}\sim C$, where C is a constant which takes different values 
depending on the case, electromagnetic, odd or even gravitational perturbations.  
So equation (\ref{waveeq}) becomes, in this limit,
\begin{equation}
Ay\frac{\partial^{2} \phi}{\partial y^{2}}+ [A-2i\omega]\frac{\partial
\phi}{\partial y} -C\phi=0\,, 
\label{waveeq1}
\end{equation}
where $y=r-r_+$, and $A=3r_+ +\frac{1}{r_+}$.  This equation has an exact
solution in terms of the modified Bessel functions $ I_{\nu}(z)$ \cite{Stegun}, 
\begin{equation}
\phi=C_1y^{i\frac{\omega}{A}}I_{-\frac{i\omega}{A}}
\left(2(\frac{C}{A}y)^{\frac{1}{2}}\right)
+C_2y^{i\frac{\omega}{A}}I_{\frac{i\omega}{A}}
\left(2(\frac{C}{A}y)^{\frac{1}{2}}\right).
\label{wavesol1}
\end{equation}
We want the asymptotic behavior of these functions when $y
\rightarrow 0$ which is given by  $I_\nu(z)
\rightarrow \frac{(\frac{z}{2})^\nu}{\Gamma(\nu +1)}\,,z \rightarrow
0$.  So, near the horizon the wavefunction $\phi$ behaves
as
\begin{equation}
\phi_{r_+}=C_1\frac{
\left(\frac{C}{A}\right)^{-\frac{i\omega}{A}}}
{\Gamma\left(1-\frac{2i\omega}{A}\right)}
+C_2\frac{y^{\frac{2i\omega}{A}}
\left(\frac{C}{A}\right)^{\frac{i\omega}{A}}}
{\Gamma\left(1+\frac{2i\omega}{A}\right)}.
\label{waveeqhor}
\end{equation}
We can see that if one wants to rule out outgoing modes at the
horizon, we must have $C_2=0$, so that $\phi$ in equation
(\ref{generic}) does not depend on $y$.  Let's now investigate the asymptotic
behavior at infinity.  For $r \rightarrow \infty$ we have
$\frac{V}{f} \rightarrow \frac{l(l+1)}{r^2}$. Therefore near infinity
equation (\ref{waveeq}) becomes
\begin{equation}
r^2\frac{\partial^{2} \phi}{\partial r^{2}}+
[2r-2i\omega]\frac{\partial \phi}{\partial r}
-\frac{l(l+1)}{r^2}\phi=0 \,.
\label{waveeq2}
\end{equation}
Putting $x=\frac{1}{r}$ we have
\begin{equation}
\frac{\partial^{2} \phi}{\partial x^{2}}+ 2i\omega\frac{\partial
\phi}{\partial x} -l(l+1)\phi=0 \,,
\label{waveeq3}
\end{equation} 
with solution $\phi=\phi_{\rm \infty}$ given by 
\begin{equation}
\phi_{\rm \infty}(x)=A \,
{\rm e}^{\left\lbrack -i\omega+i\left(\omega^2-l(l+1)\right)^{1/2}\right\rbrack x}
+ B\,
{\rm e}^{\left\lbrack-i\omega-i\left(\omega^2-l(l+1)\right)^{1/2}\right\rbrack x}
\label{wavesol2}
\end{equation}
Now, $\phi_{\rm \infty}(x=0)=0$, therefore $A=-B$, and thus,
\begin{equation}
\phi_{\infty}(x)=A\,{\rm e}^{-i\omega x}
\sin\left\lbrack \left(\omega^2-l(l+1)\right)^{\frac{1}{2}}x\right\rbrack
\label{wavesol3}
\end{equation}
We can now proceed in the proof. Multiplying equation (\ref{waveeq}) by $\bar{\phi}$ 
(the complex conjugate of $\phi$), and
integrating from $ r_+ $ to $\infty$ we obtain 
\begin{equation}
\int_{r_+}^{\infty}dr\left\lbrack\bar{\phi}
\frac{d\left(f\frac{d\phi}{dr}\right)}{dr}-
2i\omega\bar{\phi}\frac{d\phi}{dr}-\frac{V}{f}\bar{\phi}\phi \right\rbrack=0\,.
\label{1}
\end{equation}
Integrating by parts yields 
\begin{equation}
\int_{r_+}^{\infty}dr \left\lbrack
\frac{d[\bar{\phi}f\frac{d\phi}{dr}]}{dr}-
f|\frac{d\phi}{dr}|^2 - 2i\omega\bar{\phi}\frac{d\phi}{dr}-\frac{V}{f}|\phi|^2 
\right\rbrack=0\,.
\label{2}
\end{equation}
Now, one can show that $[\bar{\phi}f\frac{d\phi}{dr}]_{r_+}=0$, 
in order to satisfy the boundary conditions. Indeed,  at $r_+$, 
$\phi(r_+)={\rm constant}$ and $f(r_+)=0$.  Now, at infinity, even though
$\bar{\phi}(\infty)=0$, we have also $f(\infty)=\infty$, so we have to
show that $[\bar{\phi}f\frac{d\phi}{dr}]_{\infty}=0 $. From equation 
(\ref{wavesol3}) we can check that this is indeed true.  
Thus, equation (\ref{2}) gives 
\begin{equation}
\int_{r_+}^{\infty}dr \left\lbrack
f|\frac{d\phi}{dr}|^2 + 2i\omega\bar{\phi}\frac{d\phi}{dr}+\frac{V}{f}|\phi|^2 
\right\rbrack=0.
\label{3}
\end{equation}
Taking the imaginary part of (\ref{3}) we have
\begin{equation}
\int_{r_+}^{\infty}dr \left\lbrack
\omega\bar{\phi}\frac{d\phi}{dr}+ \bar{\omega}\phi\frac{d\bar{\phi}}{dr}
\right\rbrack=0\,,
\label{4}
\end{equation}
wich, after an integration by parts reduces to 
\begin{equation}
(\omega-\bar{\omega})\int_{r+}^{\infty}dr \left\lbrack
\bar{\phi}\frac{d\phi}{dr}
\right\rbrack
=\bar{\omega}|\phi(r_+)|^2.
\label{5}
\end{equation}
Finally, inserting this back into (\ref{3}) yields 
\begin{equation}
\int_{r_+}^{\infty}dr  \left\lbrack
f|\frac{d\phi}{dr}|^2 +\frac{V}{f}|\phi|^2 \right\rbrack
=-\frac{|\omega|^2|\phi(r_+)|^2}{{\rm Im}\, 
\omega}.
\label{6}
\end{equation}
 From this relation, one can infer that, if V is positive definite then
${\rm Im}\,\omega\,<0$ necessarily.
So, since electromagnetic and even gravitational perturbations have $V>0$ 
one always has ${\rm Im}\, \omega <0$. As for odd gravitational perturbations 
there are instances where $V<0$, making this theorem unreliable for these cases. 
However, for 
$r_+<\left\lbrack\frac{l(l+1)}{3}-1\right\rbrack^{\frac{1}{2}}$, 
i.e., small enough masses, 
$V>0$ (see equation (\ref{vodd})), and the theorem applies. 
 
Another important point concerns the late time behavior of these fields,
and the existence or not of power-law tails. As shown by Ching et al \cite{Ching}, 
for potentials that vanish exponentially near the horizon, there are no power-law tails,
so there  will be no such tails in our case.  

\subsection{Numerical Calculation of the QNM frequencies}
To find the frequencies $\omega$ that satisfy the 
boundary conditions we first note that equation (\ref{waveeq}) has only regular
singularities in the range of interest. It has therefore, by Fuchs
theorem, a polynomial solution \cite{arfken}. 
To deal with the point at infinity, we
first change the independent variable $r$  to $x=\frac{1}{r}$. Now we can use
Fr\"{o}benius method by looking for an indicial equation (for further
details see \cite{Horo}), and force it to obey the boundary condition
at the horizon ($x=\frac{1}{r_+}=h$).  We get the following solution to equation 
(\ref{waveeq}), 
\begin{eqnarray} \phi(x)=
\sum_{n=0}^{\infty} a_{n(\omega)} (x-h)^n \,,
\label{frobenius} 
\end{eqnarray}
where $a_{n(\omega)}$ is a function of the frequency.  If we put
(\ref{frobenius}) into (\ref{waveeq}) and use the boundary condition
$\phi=0$ at infinity ($x=0$) we obtain  
\begin{equation}
\sum_{n=0}^{\infty} a_{n(\omega)}(-h)^n=0 \,.
\label{numerico} 
\end{equation}
Our problem is reduced to that of finding a 
numerical solution of the polynomial equation (\ref{numerico}).
The numerical roots for $\omega$ of equation (\ref{numerico}) can be
evaluated resorting to numerical computation. Obviously, one cannot
determine the full sum in expression ($\ref{numerico}$), so we have to
determine a partial sum from $0$ to $N$, say and find the roots $\omega$
of the resulting polynomial expression.  We then move onto the next
term $N+1$ and determine the roots.  If the method is reliable, the
roots should converge. We stop our search when we have a 3 decimal digit
precision.

\subsubsection{Electromagnetic modes}
As long as the modes are decaying, it does not matter whether they're 
oscillating or not. However,
as we will see there are frequencies in the electromagnetic case 
with a vanishing real
part, which makes it possible to use an approximation, due to Liu \cite{Liu}, to
the highly damped modes. Although the method was
originally developed for the asymptotically flat space, it is quite
straightforward to apply it to our case.  There is therefore a way to
test our results. Unfortunately, this method relies heavily
on having  a frequency with a large pure imaginary part, so as we shall see
it will only work for electromagnetic perturbations.
 We have computed the lowest frequencies for some
values of the horizon radius $r_+$, and $l$.  
The frequency is written as $\omega = \omega_r +
i\omega_i$, where $\omega_r$ is the real part of the frequency and
$\omega_i$ is its imaginary part.
In tables 1 and 2 we list the numerical values of the lowest quasinormal
frequencies of electromagnetic perturbations 
for $l=1$ and $l=2$ and for selected values of $r_+$ . 
For frequencies with no real part, we list the values obtained
in Liu's aproximation.
\begin{center}
\begin{tabular}{|l|l|l|l|l|}  \hline 
\multicolumn{1}{|c|}{} &
\multicolumn{2}{c|}{ Numerical} &
\multicolumn{2}{c|}{ Liu's approximation} \\ \hline
$r_+$    &  $-\omega_i$ &  $\omega_r$ &  $-\omega_i$  & $\omega_r$ \\ \hline
0.8  & 1.287 & 2.175 & $-$ & $-$  \\ \hline
 1   &   1.699 &  2.163  & $-$ & $-$   \\ \hline
5  & 8.795 & $\sim0$ & 7.6 & $\sim0$  \\ \hline
10  &  15.506 & $\sim0$  & 15.05 & $\sim0$  \\ \hline
50  & 75.096 & $\sim0$ & 75.01 & $\sim0$  \\ \hline
100 & 150.048  & $\sim0$ & 150.005 & $\sim0$  \\ \hline
\end{tabular}
\end{center}
\vskip 1mm
{\noindent Table 1. Lowest QNM of electromagnetic perturbations for $l=1$.
The $-$ in Liu's approximation columns means that the method is not applicable.}
\vskip 8mm

\begin{center}
\begin{tabular}{|l|l|l|l|l|}  \hline 
\multicolumn{1}{|c|}{} &
\multicolumn{2}{c|}{ Numerical} &
\multicolumn{2}{c|}{ Liu's approximation} \\ \hline
$r_+$    &  $-\omega_i$ &  $\omega_r$ &  $-\omega_i$  & $\omega_r$ \\ \hline
0.8  & 1.176 & 2.501 & $-$ & $-$  \\ \hline
 1   & 1.579   &  2.496  & $-$ & $-$   \\ \hline
5  & 10.309 & 0.822 & 7.6 & $\sim0$  \\ \hline
10  & 15.755  & $\sim0$  & 15.05 & $\sim0$  \\ \hline
50  & 75.139  & $\sim0$ & 75.01 & $\sim0$  \\ \hline
100 & 150.069  & $\sim0$ & 150.005 & $\sim0$  \\ \hline
\end{tabular}
\end{center}
\vskip 1mm
{\noindent Table 2. Lowest QNM of electromagnetic perturbations for $l=2$.}
\bigskip

\noindent
As one can see, the imaginary part of the frequency, which determines
how damped the mode is, and which according to the AdS/CFT conjecture
is a measure of the characteristic time $\tau=\frac{1}{\omega_i}$ of
approach to thermal equilibrium, scales linearly (for large black
holes) with the horizon radius supporting the arguments given in
\cite{Horo}.  Moreover, the frequencies do not seem to depend on the
angular quantum number $l$, and are in excellent agreement with the
analytical approximation for strongly damped modes.

\noindent
For a better visualization we also plot $\omega_i \times r_+ $ in 
Figure 1. 
\vskip 3mm
\centerline{\epsffile{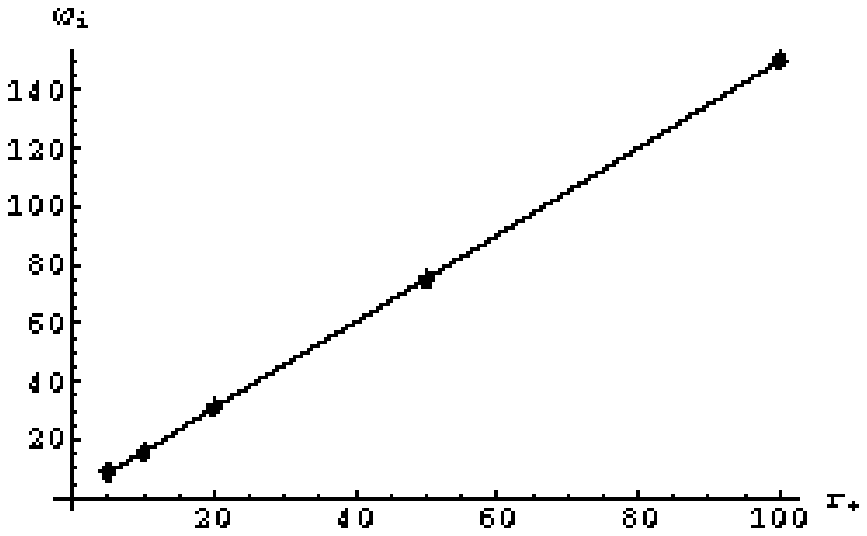}}
\vskip 1mm
{\noindent Figure 1. Lowest electromagnetic 
QNM for $l=1$ as function of $r+$. The black lozenges represent some frequencies numerically 
calculated. The line connecting them is a linear fit.}
\vskip 3mm

\subsubsection{Gravitational modes}

The numerical calculation of the quasinormal frequencies for gravitational
perturbations proceeds as outlined previously (the associated differential
equation has only regular singularities, so it is possible
to use an expansion such as (\ref{frobenius})).

\bigskip
\noindent {\bf (i) Odd modes:}
In tables 3 and 4 we show
the two lowest  QNM frequencies for $l=2$ and $l=3$ odd gravitational 
perturbations. 
An important point in odd QNMs is that there is a mode for which the
frequencies are not only pure imaginary and very small, but also scale
with $\frac{1}{r_+}$!  This is similar to the behavior of
Schwarzschild black holes in asymptotically flat spacetimes, as
mentioned.  However, all frequencies have a negative imaginary part,
which indicates that the spacetime is stable for this kind of
perturbations.

\begin{center}
\begin{tabular}{|l|l|l|l|l|}  \hline 
\multicolumn{1}{|c|}{} &
\multicolumn{2}{c|}{ lowest QNM} &
\multicolumn{2}{c|}{ second lowest QNM} \\ \hline
$r_+$    &  $-\omega_i$ &  $\omega_r$ &  $-\omega_i$  & $\omega_r$ \\ \hline
0.5   & 6.4 & 0 & 0.72 &3.037   \\ \hline
1   &$\sim$ 2(?) & 0 & 2.404 & 3.033  \\ \hline
2  & 0.728  & $\sim0$  &5.258  & 4.447   \\ \hline
5  & 0.2703 & $\sim0$ &13.294  &9.577   \\ \hline
10   &  0.13378 & $\sim0$  & 26.626 & 18.662  \\ \hline
50   & 0.02667 & $\sim0$ & 133.19 & 92.505  \\ \hline
100  & 0.0132  & $\sim0$ & 266.384 & 184.959  \\ \hline
\end{tabular}
\end{center}
\vskip 1mm
{\noindent Table 3. Lowest QNM of gravitational odd perturbations for $l=2$.}
\vskip 8mm

\begin{center}
\begin{tabular}{|l|l|l|l|l|}  \hline 
\multicolumn{1}{|c|}{} &
\multicolumn{2}{c|}{ lowest QNM} &
\multicolumn{2}{c|}{ second lowest QNM} \\ \hline
$r_+$    &  $-\omega_i$ &  $\omega_r$ &  $-\omega_i$  & $\omega_r$ \\ \hline
1   & 10 & 0 & 1.639 & 3.849  \\ \hline
2  & 2.189  & 0  &5.080  & 4.615  \\ \hline
5  & 0.690 & $\sim0$ &13.247  &9.735   \\ \hline
10   & 0.336  & $\sim0$  & 26.603 & 18.742  \\ \hline
50   & 0.0669 & $\sim0$ & 133.19 & 92.521  \\ \hline
100  & 0.0333  & $\sim0$ & 266.382 & 184.967  \\ \hline
\end{tabular}
\end{center}
\vskip 1mm
{\noindent Table 4. Lowest QNM of gravitational odd perturbations for $l=3$.}
\vskip 3mm

The value $\omega=2$ in Table 3 marked
with a ``?'' is a somewhat dubious result.  In fact, from
(\ref{waveeqhor}) it follows that if $1-\frac{2i\omega}{A}=-n$, with
$n$ an integer, then there is nothing going down the hole, so 
perhaps it is not a QNM.  It is
also an ``algebraically special value'' in the sense of Chandrasekhar
\cite{Chandrass}. 

\bigskip
\noindent {\bf (ii) Even modes:}
In table 5 we show the lowest QNM frequencies for $l=2$ and $l=3$ even
gravitational perturbations.  We point out the remarkable resemblance
of the values in table 3 with those in \cite{Horo} for scalar
perturbations, even though the potentials are quite different. We have
performed calculations for higher values of the angular quantum number
$l$, and found that the QNM frequencies are indeed very similar
throughout all values of $l$.

\begin{center}
\begin{tabular}{|l|l|l|l|l|}  \hline 
\multicolumn{1}{|c|}{} &
\multicolumn{2}{c|}{ lowest QNM, $l=2$} &
\multicolumn{2}{c|}{ lowest QNM, $l=3$} \\ \hline
$r_+$    &  $-\omega_i$ &  $\omega_r$ &  $-\omega_i$  & $\omega_r$ \\ \hline
1   &1.584 &3.018  &  1.392 & 3.909 \\ \hline
2  & 3.974  &4.546  &  3.299  & 4.597    \\ \hline
5  & 12.649 &9.83 &  11.642 &10.217    \\ \hline
10  & 26.301  &18.806  & 25.788  &19.089    \\ \hline
50  &133.125 &92.535  & 133.022  &92.596     \\ \hline
100  & 266.351  & 184.959 &  266.300  &185.005    \\ \hline
\end{tabular}
\end{center}
\vskip 1mm
{\noindent Table 5. Lowest QNM of gravitational even perturbations.}
\vskip 3mm

\bigskip
\noindent {\bf (iii) Discussion:}
We first note that there is clearly a distinction between odd
and even perturbations: they no longer have the same spectra, contrary
to the asymptotically flat space case (see \cite{Chandras}), a problem
we shall consider in more detail in the next subsection. 
 We also remark that in electromagnetic and scalar perturbations the
frequency scales with $r_+$ (for large black holes at least). Since
the temperature scales also with
 $r_+$ in the large black hole regime, this means that
 the frequency scales with the temperature.  Thus, in the dual CFT the
approach to thermal equilibrium is faster for higher temperatures.
This is a  totally different
behavior from that of asymptotically flat space, in which the
frequency scales with $\frac{1}{r_+}$.
 However, for odd modes there is one that scales 
with $\frac{1}{r_+}$. This is a reflection of the different behavior of
the potential $V_{\rm odd}$ for odd perturbations (that was why we couldn't prove
stability for odd perturbations in the first place), and of the boundary conditions, 
as we shall show in section 3.3. 
 The odd modes are therefore particularly long-lived.

\noindent
\subsection{On the isospectrality breaking between odd and even perturbations }
As is well known \cite{Chandra2,Chandras} in the case of a
Schwarzschild black hole in an asymptotically flat space the two
potentials $V_{\rm even}$ and $ V_{\rm odd}$ give rise to the same quasinormal
frequencies (in fact to the same absolute value of the reflection
and transmission coefficients).  This remarkable
property followed from a special relation (the equivalent for 
asymptotically flat spacetimes of our equation (\ref{V})) between the
potentials and the behavior of W at the boundaries.  However, as one
can see in tables 3, 4 and 5 there is a isospectrality breaking between odd and
even perturbations in Schwarzschild anti-de Sitter spacetime.

We shall now treat this problem. 
The breaking of the isospectrality is intimately
related to the behavior of $W$ at infinity.  On taking
advantage of the machinery developed by Chandrasekhar, we seek a
relation between odd and even perturbations of the form
\begin{eqnarray}
Q= p_1T+q_1\frac{dT}{dr_*} \,, \\ T= p_2Q+q_2\frac{dQ}{dr_*}\,,
\label{relation1}
\end{eqnarray}
yielding (see \cite{Chandra2} for details), 
$q_1^2=\frac{1}{\beta-\omega^2}\,$, $p_1=qW\,$, $p_2=-p_1\,$ and 
$q_2=q_1=q$. Thus, we obtain
\begin{eqnarray}
Q= qWT+q\frac{dT}{dr_*} \,, \\ T= -qWQ+q\frac{dQ}{dr_*}\,.
\label{relation2}
\end{eqnarray}
Suppose now that $\omega$ is a QNM frequency of $T$, i.e., one for which 
\begin{eqnarray}
T \rightarrow A_{\rm even} e^{-i\omega r_*} \,,\,\, r\rightarrow r_+\,, \\
T \rightarrow 0 \,,\,\, r\rightarrow \infty\,.
\label{Tasymptotic}
\end{eqnarray}
Substituting this into equation (\ref{relation2}) we see that
\begin{eqnarray}
Q \rightarrow A_{\rm even}
q\left\lbrack W(r+)-i\omega\right\rbrack 
e^{-i\omega r_*} \,,\,\,r\rightarrow r_+\,, \\
Q \rightarrow q\left(\frac{dT}{dr_*}
\right)_{r=\infty} \,,\,\, r\rightarrow \infty\,.
\label{Qasymptotic}
\end{eqnarray}
However, from equation (\ref{wavesol3}), 
$\left(\frac{dT}{dr_*}\right)_{r=\infty}$ is in general
not zero so that $\omega$ will in general fail to be a QNM frequency
for Q.  Should $Q$ and $T$ be smooth functions of $\omega$, 
one expects that if $q$
is ``almost zero'' then $\omega$ should ``almost'' be a QNM frequency
for $Q$. Now, the condition that $q$ is almost zero is that
$\beta-\omega^2$ be very large, and one expects this to be true either
when $\omega$ is very large or else when $\beta$ is very large. And in
fact, as one can see in tables 3, 4 and 5 for very large $\omega$ the
frequencies are indeed almost identical. On the other hand, for very
small black holes ($\beta$ very large) one expects the frequencies to
be exactly the same, since both potentials have the same asymptotic
behavior in this regime, as we shall see in the next section.  One
would be tempted to account for the remarkable resemblances between QNM
frequencies of scalar and gravitational perturbations by a similar
approach, but the proof is still eluding us.  Should such an approach
work, it could be of great importance not only to this specific
problem, but also to the more general problem of finding the
asymptotic distribution of eigenvalues, by studying a different
potential with (asymptotically) the same eigenvalues, but more easy to
handle.

\noindent
\section{The limit m $\rightarrow$ 0}
Although it is not possible to solve exactly for the QNM frequencies,
it is possible to gain some analytical insight in the special case of
very small black holes. There has been some discussion
about this regime (see \cite{Wang3}, and also \cite{Horo,Hubeny} 
and references therein). Here
we shall exploit the behavior of QNM frequencies in this regime a
little further.
 For very small black holes one can easily see that both
potentials (electromagnetic and gravitational) look like,
 in the $r_*$ coordinate, a barrier with
unequal heights:
\vskip 3mm
\centerline{\epsffile{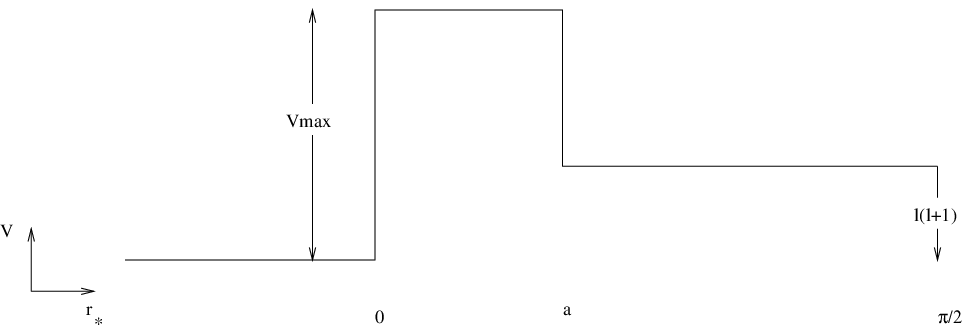}}
\vskip 1mm
{\noindent Figure 2. The potential for small black holes.
$V_{\rm max}=\frac{4l(l+1)}{27r_+^2}$ and $a \sim 3r_+$.}
\vskip 3mm
It is trivial to obtain equations for the quasinormal
frequencies in this limit. If $\Psi$ is a general wavefunction then
\begin{eqnarray}
\frac{d\Psi/dr_*}{\Psi}=-i\omega \;,\quad r_*<0\,, \\
\frac{d\Psi/dr_*}{\Psi}=\frac{ik_1Be^{ik_1r_*}-ik_1Ce^{-ik_1r_*}}
{Be^{ik_1r_*}+Ce^{-ik_1r_*}}\;, \quad 0<r_*<a\,, \\
\frac{d\Psi/dr_*}{\Psi}=\frac{ikDe^{ikr_*}-ik_1Ee^{-ikr_*}}
{De^{ikr_*}+Ee^{-ikr_*}}\;, \quad r_*>a \;,
\label{5.1}
\end{eqnarray}
where $k_1=(\omega^2-V_{max})^{1/2}$, and
$k=\left\lbrack\omega^2-l(l+1)\right\rbrack^{1/2}$. Imposing the continuity
of the logarithmic derivative and furthermore that $\Psi=0$ at
infinity ($r_*=\frac{\pi}{2}$), we get
\begin{eqnarray}
k_1\left\lbrack\frac{\frac{k_1-\omega}{k_1+\omega}e^{2ik_1a}-1}
{\frac{k_1-\omega}{k_1+\omega}e^{2ik_1a}+1}\right\rbrack
=
k\left\lbrack
\frac{e^{2ik(\frac{\pi}{2}-a)}+1}
{1-e^{2ik(\frac{\pi}{2}-a)}}\right\rbrack\;.
\label{5.2}
\end{eqnarray}
In the limit $a \rightarrow 0$, $k_1 \rightarrow \infty$
($m \rightarrow 0$) we have, supposing that $\omega$ stays small,
the condition $e^{2ik\frac{\pi}{2}}=1$ which means that
\begin{equation}
\omega_0^2=4n^2+l(l+1)\;, \quad n=1,2,...\;,
\label{5.3}
\end{equation}
corresponding to a bound state. This gives for the lowest 
QNM frequencies ($n=1$):
$\omega_0=2.45 \,$ for $l=1$ and $\omega_0=3.16 \,$ for $l=2$.
The above are to be compared with those in Tables 1-5.
The agreement seems excellent, and we can now go a step further
: If we linearize (\ref{5.2}) around the solution (\ref{5.3}), i.e,
if we write $\omega= \omega_0+i\delta$ and substitute back in (\ref{5.2}) we 
obtain, 
to third order in $\delta$, the values listed in Table 6 . 
\begin{center}
\begin{tabular}{|l|l|l|l|l|}  \hline 
\multicolumn{1}{|c|}{} &
\multicolumn{1}{c|}{$ a=3/h$} &
\multicolumn{1}{c|}{$ a=6/h$} \\ \hline
\multicolumn{1}{|c|}{l}    & \multicolumn{1}{|c|}{ $\delta$} &\multicolumn{1}{|c|}{  $\delta$}
  \\ \hline
1   &$-2.1/h-i(1.42/h^2)$  &$ -1.92/h-i(0.05/h^2)$  
\\ \hline
2  &$ -0.859/h-i(0.04/h^2) $ &
$ -0.85/h-i(1.4{\rm x}10^{-4}/h^2)$   
\\ \hline
\end{tabular}
\end{center}
\vskip 1mm
{\noindent Table 6. The linearized frequency $\delta\,$ for selected values of the 
angular quantum number $l$ and the potential width $a$. }
\vskip 3mm
We have chosen a typical value of $a\sim \frac{3}{h}$, but we can see
that, although the real part does not depend very much on $a$, the imaginary part
is strongly sensitive to $a$. Nevertheless, one can be sure that whatever
value of $a$, the imaginary part goes as $\frac{1}{h^2}$ and is always negative.

\noindent
\section{Conclusions}
 We have computed the electromagnetic and gravitational QNM
frequencies of Schwarzschild-AdS black holes in four dimensions. These
modes dictate the late time behavior of a minimally coupled
electromagnetic field and of small gravitational perturbations,
respectively.   The conclusions are:
(i) The frequencies all have a negative imaginary part, which means that
the black hole is stable against these perturbations, since these will
decay exponentially with time;
 (ii) Maxwell perturbations are strongly damped, so according
to the AdS/CFT conjecture, any electromagnetic perturbed
thermal state will rapidly approach equilibrium;
(iii) for odd gravitational perturbations in the large black hole regime, the
imaginary part of the frequency (decaying mode) goes to zero scaling with
$\frac{1}{r_+}$, just as in asymptotically flat space.  In terms of the
AdS/CFT correspondence, this implies that the greater the mass, the
more time it takes to approach equilibrium, an unusual result; 
(iv) scalar \cite{Horo} and gravitational even perturbations exhibit an amazing
similarity for the characteristic time damping of the perturbations,
but we have not been able to prove it analytically;
 (v) in the small black hole regime
the imaginary part of the frequency (decaying mode) scales with $r_+^2$.

\vskip 3mm

\vskip .5cm

\section*{Acknowledgments} 
This work was partially funded
by Funda\c c\~ao para a  Ci\^encia e Tecnologia (FCT) 
through project PESO/PRO/2000/4014. V.C.  also 
acknowledges finantial support from FCT 
through PRAXIS XXI programme.
J. P. S. L. thanks Observat\'orio Nacional do Rio de Janeiro for
hospitality.

\vskip .5cm
\newpage


\end{document}